\documentclass[pre,aps,twocolumn,showpcs,floatfix,amsmath,amssymb]{revtex4}
\usepackage{epsfig}
\usepackage{amsmath}

\def\qbold{\mbox{\boldmath $q$\unboldmath}}
\def\Dbold{\mbox{\boldmath $D$\unboldmath}}
\def\Fbold{\mbox{\boldmath $F$\unboldmath}}
\def\rbold{\mbox{\boldmath $r$\unboldmath}}

\def\ubold{\mbox{\boldmath $u$\unboldmath}}

\def\3dots{\:\raisebox{-0.5ex}{$\stackrel{\textstyle.}{:}$}\:}
\def\beq{\begin{equation}}
\def\eeq{\end{equation}}
\def\bea{\begin{eqnarray}}
\def\eea{\end{eqnarray}}
\begin{document}
\title{The mechanics of anisotropic spring networks}
\author{T. Zhang$^1$, J. M. Schwarz$^1$, and Moumita Das$^2$}
\affiliation{$^1$Department of Physics, Syracuse University, Syracuse, NY 13244, $^2$School of Physics and Astronomy, Rochester Institute of Technology, Rochester, NY 14623, USA}
\begin{abstract}
We construct and analyze a model for a disordered linear spring network with anisotropy. The modeling is motivated by, for example, granular systems, nematic elastomers, and ultimately cytoskeletal networks exhibiting some underlying anisotropy. The model consists of a triangular lattice with two different bond occupation probabilities, $p_x$ and $p_y$, for the linear springs. We develop an effective medium theory (EMT) to describe the network elasticity as a function of $p_x$ and $p_y$. We find that the onset of rigidity in the EMT agrees with Maxwell constraint counting.  We also find beyond linear behavior in the shear and bulk modulus as a function of occupation probability in the rigid phase for small strains, which differs from the isotropic case. We compare our EMT with numerical simulations to find rather good agreement. Finally, we discuss the implications of extending the reach of effective medium theory as well as draw connections with prior work on both anisotropic and isotropic spring networks.  
\end{abstract}
\maketitle
\section{Introduction}
The onset of rigidity in disordered spring networks has been studied as a model for elasticity in disordered materials. Such a model undergoes a phase transition from not-rigid to rigid at some critical fraction of springs (bonds)~\cite{degennes,feng,thorpe,duxbury}. Numerical simulations on a bond-diluted triangular lattice suggest that the transition is a continuous one with the percolating rigid cluster having a fractal dimension at the transition~\cite{thorpe,thorpe2}. To date, one of the few theoretical tools to analyze rigidity percolation is effective medium theory (EMT). In 1985, an effective medium theory was developed by Feng, Thorpe, and Garboczi analytically capturing the relationship between disorder and mechanical response in these systems in good agreement with numerical simulations~\cite{feng1, feng2, thorpe.cai}. 

We now extend this theoretical framework -- the EMT and the numerical simulations -- to a disordered spring network with anisotropy.  Anisotropic networks abound in nature and man-made materials, and are an important class of soft matter.  Consider, for example, an orientationally ordered, but spatially disordered, network made of liquid crystal elastomers. The interplay between the gel-sol transition and the isotropic-nematic transition in both synthetic and biological systems has been studied~\cite{terentjev,dalhaimer}. Consider also a layered, but disordered, system of granular particles~\cite{granular1}.  What are the macroscopic elastic properties of such a system? Some properties have been analyzed in the ordered case~\cite{goldenberg,otto}, but the disordered case is more complex and has received less attention.  Finally, the cytoskeleton, the filamentous scaffolding that provides most animal cells their shape and rigidity, can consist of rather oriented, cross-linked filaments such as in lamellipodia, the broad, thin protrusion at the leading edge of a crawling cell~\cite{svitkina,blanchoin}. 

While this paper focuses on extending the analytical framework of effective medium theory to disordered linear spring networks with anisotropy, the rigidity transition in ``super-elastic'' anisotropic central-force networks has been studied in prior work by Roux and Hansen and Wang and Harris~\cite{hansen,harris1,harris2}. In these ``super-elastic'' networks, all bonds in a preferred direction are occupied with springs, and bonds in other directions have infinite rigidity with a probability $p$ and a finite rigidity with a probability $1-p$. In addition to determining the rigidity percolation threshold, the notion of splay rigidity, in which only rotational degrees of freedom are frozen out, and a mapping to a random resistor network for the bulk modulus is discussed~\cite{harris1}. Some of their results can be extended to the usual bond-diluted system that we study here. However, there remain open questions about the interplay between anisotropy and rigidity in these systems. In particular, one can ask how does the difference in the directional occupation probability of springs influence the mechanical response of the network, and how does this couple to the direction of the applied deformation? 

To answer these questions, we investigate a triangular lattice based anisotropic bond-diluted network and study how the anisotropy in the occupation of the springs influences the ability of the network to bear stresses using an effective medium theory and numerical simulations.  The manuscript is set up as follows. We first describe the model network and its constitutive properties in Section II. This is followed by Section III with the analysis of the model using a Maxwell constraint counting argument, the description of an effective medium theory, and a conjugate gradient numerical minimization approach that we use to further investigate the model. In Section IV, we present and discuss our results, and in Section V, we comment on their implications for the relevant systems at hand.

\section{Model}

We begin with a fully ordered, but anisotropic network of springs arranged in a two-dimensional triangular lattice. The bonds are given an extensional spring constant $\alpha$ for springs in the $x$ direction, and $\gamma$ for springs making $60^\circ$ and $120^\circ$ angle to the $x$ direction, i.e. having a $y$ component. We then introduce disorder into the system by removing bonds along the $x$ direction with probability $1-p_x$, where $0 < p_x < 1 $, and bonds with a $y$ component with probability $1-p_y$, where $0<p_y<1$. There are no spatial correlations between these cutting points in either case. This generates a disordered network with a broad distribution of spring lengths in either direction.  When two springs intersect, there exists a cross-link preventing the two springs from sliding with respect to one another, but they can rotate freely without any energy cost. 

\begin{figure}
\begin{center}
\includegraphics[width=7.5cm]{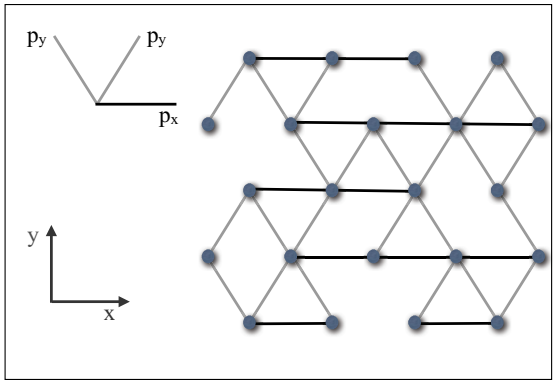}
\caption{(Color Online) (Color Online) Schematic figure showing the randomly diluted anisotropic spring network with corresponding occupation probabilities $p_x$ and $p_y$.}
\label{model}
\end{center}
\end{figure}

We study the mechanical response of this disordered network under an externally applied strain in the linear response regime. 
For simplicity we set the rest length of the springs to unity. Let  $\rbold_{ij}$ be the unit vector along the spring $ij$ and $\ubold_{ij}=\ubold_{i}-\ubold_{j}$ be the deformation of this spring. 
For small deformations, the deformation energy can be written as follows:

\bea \label{energies}
E  =  \frac{\alpha}{2} \sum_{\langle ij \rangle}\! p_{x,ij}  \left (\ubold_{\alpha,ij} . \rbold_{\alpha,ij} \right)^2+ \frac{\gamma}{2} \sum_{\langle ij \rangle}\! p_{y,ij}  \left (\ubold_{\gamma,ij} . \rbold_{\gamma,ij} \right)^2,
\eea
where $p_{x,ij}$ ($p_{y,ij}$) is the probability that the $ij$ bond in the $x$ ($y$) direction is occupied as shown in Fig.\ref{model} and $a$ is the lattice spacing and is set to $1$. The deformation energy corresponds to the cost of extension or compression of the springs. Although the model allows for anisotropy in disorder as well as in constitutive elasticity of the springs, we have set the bare elastic constant of both types of springs to have the same value ($\alpha=\gamma=1$ in arbitrary units). We investigate the shear and bulk moduli of this disordered network as a function of the direction dependent occupation probability of springs in response to suitable strains imposed on the boundaries.

\section{Methods and Analysis}
\subsection{Constraint counting argument and the rigidity threshold}

We start with a constraint counting argument due to Maxwell \cite{maxwell,alexander,feng1}, a very powerful and simple way to estimate at what occupation probability the phase transition takes place. Consider a $d$ dimensional system with $N$ particles or points, and hence $N d$ degrees of freedom. The number of zero-frequency modes $(\phi Nd)$, where $0<\phi\le 1$, is equal to the number of degrees of freedom $(Nd)$ minus the number of constraints, which in this case is $(\frac{1}{2}z_{x}Np_{x}+\frac{1}{2}z_{y}Np_{y})$, where $z_{x}$ is the number of nearest-neighbor points in the $x$ direction and  $z_{y}$ is the number of nearest-neighbor points in the $y$ direction. Here, $z_{x}=2$ and $z_{y}=4$. Hence, the fraction of zero-frequency modes is 
\beq 
\phi=1-(\frac{1}{2}p_{x}+p_{y}). \nonumber \\
\eeq
So the transition takes place when $\phi$ goes to zero or 
 \beq
 \frac{p_{x}}{2}+p_{y}=1. \nonumber \\
 \eeq

\begin{figure}
\begin{center}
\includegraphics[width=8cm]{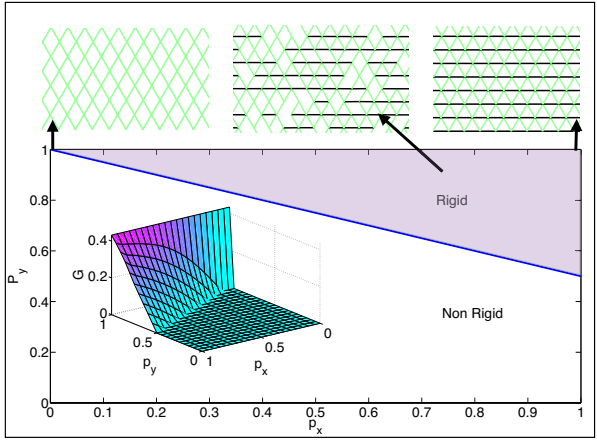}
\caption{(Color Online) (Color Online) Plot of the phase diagram according to mean field constraint counting argument, with the inset showing the shear modulus $G$ as a function of $p_x$ and $p_y$ obtained from the EMT in units of $\gamma$ (set to unity) and the horizontal lattice spacing (set to unity). Different lattice realizations are also shown. } 
\label{Figure2}
\end{center}
\end{figure}

In Fig.~\ref{Figure2}, we show the rigidity phase diagram of the disordered network, as a function of the occupation probability $p_x$ and $p_y$ of springs in the $x$ and $60^\circ$ and $120^\circ$ to the $x$ direction, respectively. 
We also show the network structure for three representative points in the rigid phase of the phase diagram: $p_x=1,\,p_y=1$, $p_x=0.75,\,p_y=0.90$, and $p_x=0,\,p_y=1$. In what follows we investigate how the mechanical response of the system changes as the network is progressively diluted, finally reaching the transition threshold. To accomplish this objective, we have used an effective medium theory and an energy minimization approach, which we describe below.    

\subsection{Effective Medium Theory}

 We study the mechanical response of this disordered network for small deformations using an effective medium theory~\cite{feng,feng1,feng2,thorpe.cai,garboczi}. 
The aim of the theory is to construct an effective medium or ordered network that has the same mechanical response as the depleted network under consideration. 
The effective filament stretching elastic constants are determined by requiring that strain fluctuations produced in the original, ordered network by randomly cutting 
filaments have zero average.

We first illustrate how the effective medium elastic constant can be calculated  for the simple case where we apply a uniform strain on an isotropic central force network, so that all bonds are equally stretched by an amount ${\delta \ell}_m$ with effective medium spring constant $\alpha_m$. 
 Let us now replace a spring between two points, say,  $i$ and $j$  by different one with spring constant $\alpha$. It would lead to additional extension or compression of this spring, which we calculate as follows~\cite{feng1,feng2}.

The virtual force necessary to return $i$ and $j$ to their original positions before the replacement of the spring is $f={\delta \ell}_m (\alpha_m - \alpha)$. If this force is now applied between $i$ and $j$ in the unstrained and ordered network, it will lead to a deformation $\delta u$ of this spring given by $f/\delta u = \alpha_m/a^*$. The effective spring constant
$\alpha_{eff}=\alpha_m/a^*$ takes into account,  through the dynamical matrix, the elasticity of the entire network  including the direct connections between these points.  
If the force $f$ is applied now on an unstrained network where the spring between $i$ and $j$ has been replaced by a spring with stretching constant $\alpha$, $f/\delta u= \alpha_{eff} = \alpha_m/a^{*} -\alpha_m + \alpha$. Therefore  change or fluctuation $\delta u$ of the bond between $i$ and $j$  is given by
\beq
\delta u = {\delta \ell}_m  \frac{\alpha_m - \alpha}{\alpha_m/a^{*} -\alpha_m + \alpha}.
\eeq

It follows from the superposition principle that this fluctuation $\delta u$ is the same as the extra extension or compression in the strained network due to the replacement of the bond $ij$.
Considering random bond dilution in the network and defining an effective medium such that the fluctuations $\delta u$ should vanish when averaged over the entire network,
\beq
\langle \delta u \rangle = 0.
\eeq
For a distribution of bonds $P(\alpha')= p \delta(\alpha'-\alpha) + (1-p) \delta(\alpha')$, with $p$ being the probability that a bond is present,  the effective medium spring constant $\alpha_m$ is given by:
\beq
\frac{\alpha_m}{\alpha}= \frac{p-a^{*}}{1-a^{*}},
\eeq
with $a^{*} = \frac{2}{N z} \sum_{q} Tr  \left[ \Dbold (q) {\Dbold}^{-1} (q) \right ] =  2/3$ for a central force network \cite{feng1}.

Now let us consider an anisotropic network where the spring constant and probability of occupation for springs in the $x$ direction ($\alpha$,\,$p_x$) and springs making $60^\circ$ and $120^\circ$ angles with the $x$ direction ($\gamma$,\,$p_y$) are different as described in Eq.~\ref{energies}. We decompose the triangular lattice system into two interconnected subsystems as shown in the schematic (Fig.\ref{model}) and calculate the stretching forces and strain fluctuations $\ubold_\alpha$ and $\ubold_\gamma$ for these two subsystems separately.  For small deformations, the restoring forces on the springs are given by:
\begin{eqnarray}
\Fbold_{\alpha,ij} &=& \alpha_m \sum \ubold_{\alpha,ij} \cdot \rbold_{\alpha,jk} \; \rbold_{\alpha,jk} \nonumber \\
\Fbold_{\gamma,ij} &=& \gamma_m \sum \ubold_{\gamma,ij}  \cdot \rbold_{\gamma,jk} \; \rbold_{\gamma,jk}
\end{eqnarray}

The $x$ and $y$ components of the deformation can be written as  $\ubold_{\alpha,\gamma} (q) =  - \Dbold^{-1} (q) \Fbold_{\alpha,\gamma} (q)$, where $\Dbold(q)$ is the dynamical matrix \cite{feng1} of the fully ordered lattice. Following the procedure for the isotropic network discussed above, we can calculate strain fluctuations in the depleted network, and effective medium elastic constants $\alpha_m$ and $\gamma_m$ by demanding that the strain fluctuations vanish when averaged over the entire network.  Since we consider uncorrelated distributions of the elastic constants $\alpha$ and $\gamma$, the effective medium elastic moduli $\alpha_m$ and $\gamma_m$ are given by
\begin{eqnarray}
\label{Triangular-EMT-Rp}
\frac{\alpha_m}{\alpha} &=& \frac{p_x - a^{*}}{1-a^{*}} \nonumber \\
\frac{\gamma_m}{\gamma} &=& \frac{p_y-b^{*}}{1-b^{*}}, 
\end{eqnarray}
above the rigidity percolation threshold, and $\alpha_m=\gamma_m=0$ below, with $p_x$ and $p_y$ at the threshold obeying the constraint condition $p_y + p_x/2 =1$ as discussed earlier. 
The geometric constants $a^{*}$ and $b^{*}$ represent the whole network contribution to the effective spring constants 
 $\alpha_m/a^{*}$ and $\gamma_m/b^{*}$ of the bonds. When the network is strained springs in the $y$ direction will contribute to the elasticity in the $x$ direction
 and vice versa due to the coupling between the two-sublattices accounted for by $a^{*}$ and $b^{*}$.
They are given by:
\bea
 a^{*} &=& \frac{2 \alpha_m}{Nz_{\alpha}} \sum_{q} Tr  \left[ (1 - e^{-i a \qbold.\rbold_{\alpha,ij}} )  \rbold_{\alpha,ij}  \rbold_{\alpha,ij} \Dbold^{-1} (q) \right ]  \nonumber \\
 b^{*} &=& \frac{2 \gamma_m}{Nz_{\gamma}} \sum_{q} Tr  \left[ ( 1 - e^{-i a \qbold.\rbold_{\gamma,ij}} )  \rbold_{\gamma,ij}  \rbold_{\gamma,ij}  \Dbold^{-1} (q) \right ]. 
 \label{astarbstar}
\eea
The sum is over the first Brillouin zone and $z_{\alpha,\gamma}$ are coordination numbers, and $\rbold_{\alpha}$ are unit vectors along bonds with spring constants $\alpha$, 
i.e. bond that were originally along the $x$ direction, while $\rbold_{\alpha,\gamma}$ are unit vectors  along bonds with spring constants $\alpha, \gamma$ in the original undeformed lattice. 
The above definition of $a^{*}$ and $b^{*}$, together with fact that in the fully ordered triangular lattice network, $1/3$ of all the bonds 
have a spring constant $\alpha$ and $2/3$ of the bonds by spring constants $\gamma$ lead to the constraint condition: $\frac{1}{3} a^{*} +\frac{2}{3} b^{*}=\frac{2}{3}$. 
At the rigidity percolation threshold, $a^{*}=p_x$ and $b^{*}=p_y$, and thus $\frac{p_x}{2} +  p_y = 1$, in agreement with the Maxwell constraint counting.
We obtain the effective medium spring constants $\alpha_m$ and $\gamma_m$ by solving equations (\ref{Triangular-EMT-Rp}) and (\ref{astarbstar}) simultaneously. These elastic constants describe an ordered network
that has the same mechanical response as the original depleted network, and can be used to calculate the shear and bulk moduli of the latter as discussed in Section IV.   

\subsection{Numerical Simulations}

Simulations are conducted on the triangular lattice with system size
$N_{x}=N_{y}=128$ (shown unless otherwise specified). The network
is initialized by adding bonds in the x direction with probability
$p_{x}$, and the bonds with a y component with probability $p_{y}$. 

 Since the model system is anisotropic, there exist more than one shear modulus.
We calculate the following shear response: a shear deformation is applied to two fixed boundaries 
along the $x$ direction with strain magnitude $\epsilon=1\%$, while the other two boundaries 
are periodic. Our simulations suggest that the mechanical response of the network may be reasonably 
approximated by linear response at such strains. Then the network is relaxed by minimizing the total energy of this
system using the conjugate gradient method~\cite{numrecipes} allowing the deformation
to be fully propagated. Eventually a minimum energy state is be
found within the tolerance $10^{-8}$ with energy $E_{min}$. Then
the shear modulus is calculated by $G=\frac{2E_{min}}{A_{unit}N_{x}N_{y}\epsilon^{2}},$where
$A_{unit}=\frac{\sqrt{3}}{2}$ , denoting the area of one unit cell
with unit bond length. Sample averaging is performed over 10 runs
typically. 

For the measurement of bulk modulus, a small ($\epsilon=2\%$) uniform strain  is applied to
all four fixed boundaries. Once
the system energy is minimized, we calculate the energy $E_{box}$
from part of the network within a box in the center of the system
with box size $N_{x}'=N_{y}'=108$. The bulk modulus is then calculated
by $K=\frac{2E_{box}}{a_{unit}N_{x}'N_{y}'\epsilon^{2}}$, where $A_{unit}=\frac{\sqrt{3}}{2}$.

\section{Results}

We now discuss our main results on the mechanical response of the anisotropic disordered network under the shear
and hydrostatic strains. To compare the results of the simulation on the shear modulus $G$ 
and bulk modulus $K$ of the network  with the effective medium theory, we first calculate the corresponding effective medium predictions 
in terms of the previously calculated spring constants $\alpha_m$ and $\gamma_m$. To do so, consider a fully ordered triangular network with central force interactions only. For small, uniform deformations, the strain energy density of a unit hexagonal cell is given by
\begin{equation}
F= \frac{1}{4 \sqrt{3}} \Sigma_{b=1}^{6}  \alpha^{(b)} n^{(b)}_i n^{(b)}_j
n^{(b)}_k n^{(b)}_m u_{ij} u_{km},
\end{equation}
where the superscript $b=1,2,3,4,5,6$ represents the six nearest neighbor bonds in the unit cell and $\alpha^{(b)}$ is the corresponding spring constant. 

For the system under study, the unit vectors $n^{(b)}$ and respective angles $\theta^{(b)}$  for the springs with spring constant $\alpha$ are given by
\begin{eqnarray}
\theta^{(1)}&=&0, n^{(1)}_1=1, n^{(1)}_2=0\\ \nonumber
\theta^{(2)}&=&\pi, n^{(2)}_1=-1, n^{(2)}_2=0 \\ \nonumber
\end{eqnarray}

Similarly, for springs with $\gamma_m$,
\begin{eqnarray}
\theta^{(3)}&=&\frac{\pi}{3}, n^{(3)}_1=\frac{1}{2}, n^{(3)}_2=\frac{\sqrt{3}}{2} \\ \nonumber
\theta^{(4)}&=&\frac{2 \pi}{3}, n^{(4)}_1=\frac{-1}{2}, n^{(4)}_1=\frac{\sqrt{3}}{2} \\ \nonumber
\theta^{(5)}&=&\frac{\pi}{3}, n^{(5)}_1=\frac{1}{2}, n^{(5)}_2=\frac{-\sqrt{3}}{2} \\ \nonumber
\theta^{(6)}&=&\frac{2 \pi}{3}, n^{(6)}_1=\frac{-1}{2}, n^{(6)}_2=\frac{-\sqrt{3}}{2} \\ \nonumber
\end{eqnarray}

With these inputs, the deformation energy density $F$ of the effective medium anisotropic network is given by~\cite{otto,micromechanics}:
\beq
\begin{split}
F=&\frac{1}{8 \sqrt{3}} [ (8 \alpha_m + \gamma_m) {u_{xx}}^2 + 9 \gamma_m {u_{yy}}^2  \\
 &+ 6 \gamma_m u_{xx} u_{yy} +  3 \gamma_m (u_{xy} + u_{yx})^2 ] \\
\end{split}
\label{def-energy}
\eeq
The stress components can then be calculated using $\sigma_{ij}= \frac{\partial F}{\partial u_{ij}}$, and can be used to calculate the direction dependent shear and bulk moduli as shown below~\cite{behroozi}. 

In the simulation we calculate the shear modulus corresponding to the boundary applied shear strain $u_{xy}$, and the 2D bulk modulus corresponding to a hydrostatic 
compression. The shear modulus for a shear strain applied via the boundaries along the $x$ direction at $L_y=0$ and $L_y=\frac{\sqrt{3}}{2} \left( N_y-1\right)$ (in units of the lattice spacing) is given by 
\beq
G=\frac{\sigma_{xy}}{u_{xy}}  =  \frac{\sqrt{3}}{4} \gamma_m.
\eeq
Under the hydrostatic compression of the system, the network undergoes a uniform compression by an amount $\delta$ in both $x$ and $y$ directions.
Clearly $u_{xx}=u_{yy}=\delta$ and $u_{xy}=0$. The area bulk modulus is given by 
\beq
K= \frac{\sigma_K}{\Delta A/A},
\eeq
where the hydrostatic stress $\sigma_K$ is given by $\sigma_K= \frac{\sigma_{xx} + \sigma_{yy}}{2} =\frac{(\alpha_m + 2 \gamma_m)\delta}{\sqrt{3}}$, and the change in area of the system relative to its original area is given by $\Delta A/A = u_{xx} + u_{yy} = 2 \delta$. The area bulk modulus is then $K= \frac{\alpha_m+ 2 \gamma_m}{\sqrt{12}}$. Note that, we recover the expected results $G=\frac{\sqrt{3}}{4} \gamma_m$ and $K=\frac{\sqrt{3}}{2} \gamma_m$ for the isotropic case. 

\begin{figure}
\begin{center}
\includegraphics[width=10cm]{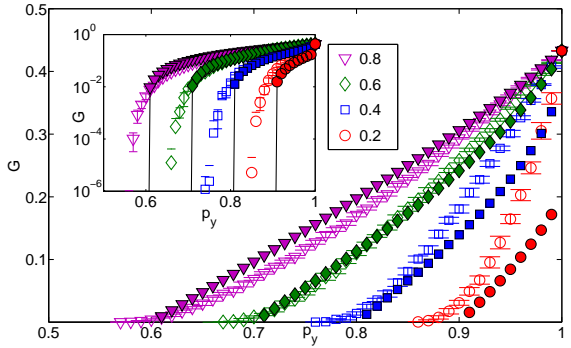}
\caption{(Color Online) The shear modulus $G$ as a function of  $p_y$ for different $p_x$ as shown in the legend. The open symbols show data from the simulations and the filled symbols (joined by solid lines in the inset) represent the result from the effective medium theory. The inset shows the same data on a log-linear scale. The system size in the simulation is $N_x=N_y=128$. }
\label{Figure3}
\end{center}
\end{figure}

Figure \ref{Figure3} shows the shear modulus obtained from the numerical simulation (open symbols) against the effective medium theory (solid lines). 
We keep $p_x$ fixed at different values, and study how $G$ changes as a function of $p_y$. We find that the agreement on the value of the shear moduli between the theory and simulation to be rather good for larger values of $p_x$ and $p_y$. In addition, our effective medium calculations suggest that for the network to have finite rigidity, $\frac{p_x}{2} + p_y  \ge 1$, i.e. it is a necessary but not sufficient condition. Random dilution of the triangular lattice leads redundant bonds and floppy inclusions being introduced~\cite{thorpe2,alexander}. Such beyond mean field effects are not taken into account in the Maxwell constraint counting.  In other words, the lattice is not cleverly constructed so that, at the transition for example, the network is minimally rigid. One ultimately needs to go beyond mean field (or EMT) and take into account the spatial makeup of the network to determine the precise value of the threshold.  Also, the subtraction of the global degrees of freedom are not included in the above condition. 

We observe that the numerically calculated value of the threshold $p_y$ is less than the analytical estimate. In addition to redundant bonds, floppy inclusions, and subtracting the global degrees of freedom, another reason for the discrepancy is that in the simulation there are two boundaries fixed where the shear deformation is applied. For the vertices on these fixed boundaries, they lose some neighbors compared to those with periodic boundary conditions, but they will add to the number of constraints since they are fixed. This would suggest a smaller threshold of $p_y$ for a given $p_x$ in the simulations as seen. 

\begin{figure}
\begin{center}
\includegraphics[width=10cm]{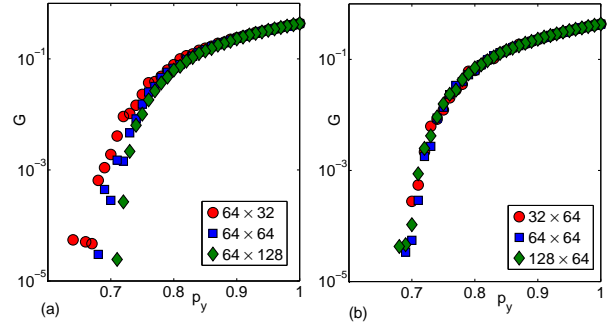}
\caption{(Color Online) Shear modulus G (for shear applied via the $x$ boundaries) as a function of $p_y$ at fixed $p_x=0.5$ and for different system sizes
(as shown in legend).}
\label{Figure4}
\end{center}
\end{figure}

To begin to quantify such boundary effects we study the dependence of the rigidity percolation threshold on system size as shown in Fig.~\ref{Figure4}. We find that while the numerically calculated value of the threshold is always
less than the analytical estimate, it moves closer to the analytical value with increasing system size. Our results suggest that the system-size dependence in the $y$ direction is stronger than in the $x$ direction.
This may be because the x-direction has periodic boundary conditions, while in the $y$ direction the boundary conditions are fixed. For a given size, systems with periodic boundaries tend to be less sensitive to finite system size effects as compared to those with fixed boundaries. We must also point out that the finite system size effects are even more pronounced the larger the difference between $p_x$ and $p_y$.  

\begin{figure}
\begin{center}
\includegraphics[width=10cm]{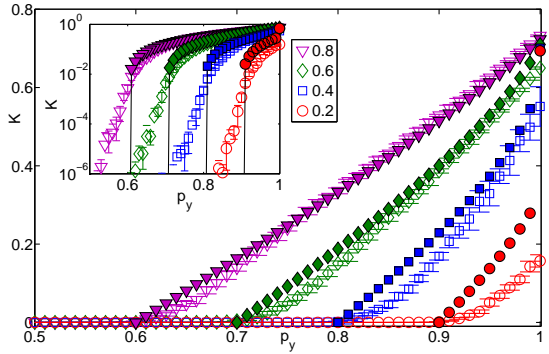}
\caption{(Color Online) The area bulk modulus $K$ as a function of $p_y$ for different $p_x$ as shown in the legend in units of the bare spring constant $\gamma$ (set to unity). The open symbols show data from the simulations and the filled symbols (joined by solid lines in the inset) represent the results from the effective medium theory. The system size in the simulation is set to $N_x=N_y=128$ and the modulus is calculated from part of the network within a box in the center of the system with box size $N_{x}'=N_{y}'=108$.}
\label{Figure5}
\end{center}
\end{figure}

\begin{figure}
\begin{center}
\includegraphics[width=10cm]{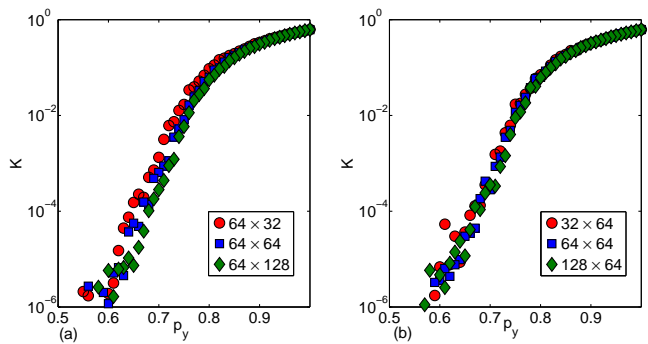}
\caption{(Color Online) Area bulk modulus K  as a function of $p_y$ at fixed  $p_x=0.5$ and for different system sizes (as shown in legend).}
\label{Figure6}
\end{center}
\end{figure}

Finally, we also study the bulk modulus $K$ in response to a hydrostatic compression by up to $2\%$. Our results are shown in Fig.~\ref{Figure5}.  As with the shear modulus, we find the best match between
the simulations and the analytical estimates on the modulus for larger values of $p_x$ and $p_y$. We also find that the numerically calculated value of the threshold is 
less than that the analytical estimate, and moves closer to the analytical value with increasing system size. The system size analysis for the bulk moduli as shown in Fig.~\ref{Figure6}, once again, shows specifically the rigidity percolation threshold
moving closer to the theoretical prediction with increasing system size and that changes in the system size in the $y$ direction lead to larger shift in the moduli as compared to changes in the system size in the $x$ direction.

Before concluding, let us discuss two limits that closely relate to prior work. The first is $p_x=1$. Wang and Harris study an anisotropic spring network on a triangular lattice where $p_x=1$ and $0<p_y<1$~\cite{harris1}. They propose the existence of a splay rigid phase in which the rotational degrees of freedom potentially freeze out at a smaller occupation probability than the translational degrees of freedom.  For the isotropic case, rotational and translational degrees of freedom become constrained across the system at the same occupation probability. When $p_x=1$, a splay rigid phase exists for $p_y>0$. It seems, however, that the onset of splay rigidity coincides with rigidity, which also coincides with the connectivity bond percolation threshold on the square lattice, which is $p_c=1/2$. Our effective medium theory predicts a rigidity percolation threshold $p_{y,rp}=1/2$ with our lattice simulations yielding $p_{y,rp}\approx 0.4$ (for our largest system size). See Fig. 7.  

When $p_x=1$, Wang and Harris, following Roux and Hansen, consider the dual anisotropic problem of a spring with infinite spring constant with probability $p_y$ and finite spring constant for probability $1-p_y$, otherwise known as the ``super-elastic'' case~\cite{hansen,harris1}. Wang and Harris demonstrate that the behavior of the bulk modulus $K$ should be identical to the conductance exponent in random resistor networks in the super-elastic network case~\cite{harris1}.  Random resistor networks are scalar analogues to the vectorial force/rigidity percolation~\cite{randomresistor}. It is not obvious whether their results can be extended to the bond-diluted case studied here. And, in fact, our analytical results suggest that at $p_{y,rp}=1/2$, $K$ is has a finite value proportional to the spring constant $\alpha$ of the springs in the $x$ direction, i.e $K$ jumps discontinuously from zero for $p_y<1/2$ (See Fig. 7). The simulations show a $K$ increasing from $\sim 0$ in not as dramatic way due to the finite size of the system. More detailed finite system size studies are needed to determine the existence of a jump in the lattice simulations. The shear modulus $G$, on the other hand, increases continuously from zero as a function of $p_y$.  

\begin{figure}
\begin{center}
\includegraphics[width=9cm]{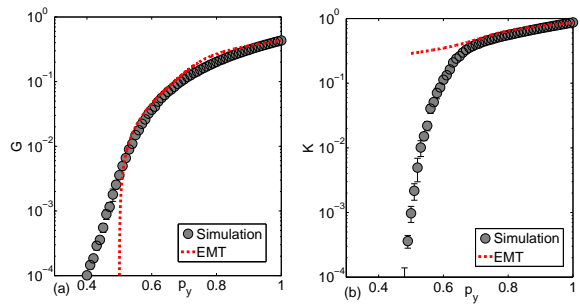}
\caption{(Color Online) The shear modulus $G$ and bulk modulus $K$ as a function of $p_y$ at fixed $p_x=1$.}
\label{Figure7}
\end{center}
\end{figure}

As for the second limit, the system is isostatic when $p_y=1$ and $p_x=0$ and periodic boundary conditions are implemented.  For fixed boundary conditions, the system is hyperstatic (over-constrained) and for free boundary conditions, the system is hypostatic (under-constrained). In the periodic case, we expect a rigidity transition as soon as $p_x>0$ as dictated by Maxwell constraint counting. This expectation is also related to work by Mao and collaborators beginning with a fully occupied square lattice of springs and adding next-nearest-neighbor springs with probability $p_{NNN}$~\cite{mao1,mao2}. In this model, the system is rigid for $p_{NNN}>0$, i.e. the transition occurs at $p_{NNN}=0$. The result goes beyond the mean-field Maxwell constraint counting, which, again, is a necessary, but not sufficient condition for rigidity~\cite{mao2}. In addition, $G$ becomes non-zero continuously with $G\sim p_{NNN}^2$~\cite{mao1}. In our model, $p_x$ is the corresponding $p_{NNN}$, however, it is a nearest neighbor bond.  Interestingly, we obtain a discontinuous onset in $G$ at $p_x=0$, where the rigidity transition occurs. See Fig. 8. This is because our shear is applied 45 degrees to the $p_y$ bonds with fixed boundary conditions in the $y$-direction, while in the earlier work shear is applied perpendicularly to the vertical square lattice bonds. Given the relation between the effective spring constants and $K$, $K>0$ as well at the transition.  As $p_x$ increases above zero, $G$ remains constant while $K$ increases to its respective limiting value. We find that $\alpha_m(p_x)-\alpha_m(p_x=0)$ increases linearly with $p_x$ to compare with the quadratic behavior found in the square lattice with additional next-nearest neighbor bonds. See Fig. 8. 

\begin{figure}
\begin{center}
\includegraphics[width=9cm]{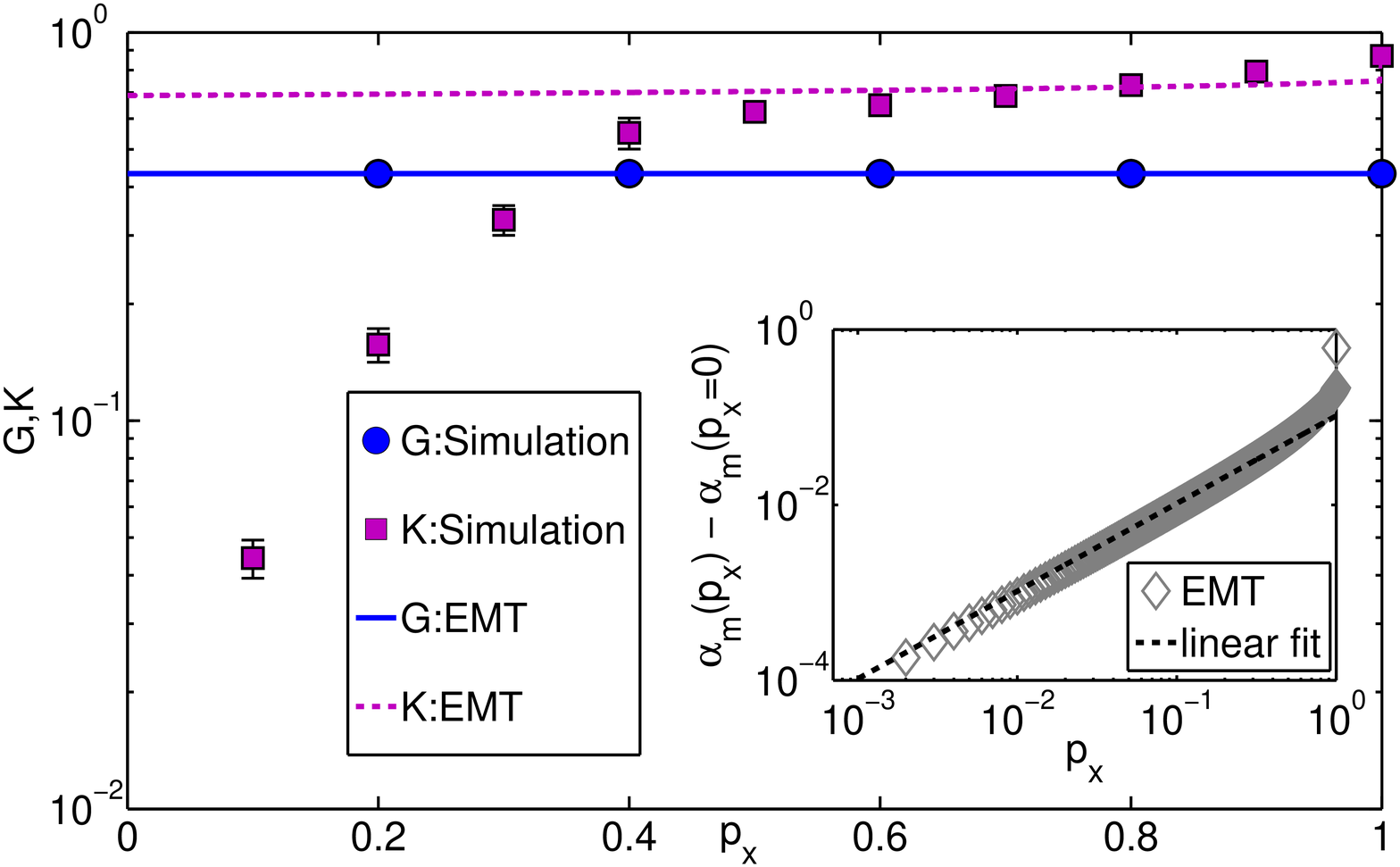}
\caption{(Color Online) The shear modulus $G$ and bulk modulus $K$ as a function of $p_x$ at fixed  $p_y=1$. The inset shows the effective medium elastic constant $\alpha_m$ as a function of $p_x$ at fixed $p_y=1$.}
\label{Figure8}
\end{center}
\end{figure}

\section{Discussion}

We have now extended effective medium theory (EMT) to anisotropic disordered spring networks. We have done so by considering a particular type of anisotropy in the occupation probability such that the triangular lattice can be considered as two interconnected sub-lattices leading to two coupled equations for the effective medium elastic constants, $\alpha_m$ and $\gamma_m$.  The elastic response of the system, such as the bulk and shear moduli, depend on these effective medium constants. For example, the shear modulus $G_{xy}$ was found to $\sqrt{3} \gamma_m /4$ with both the effective medium theory and simulations. Given the anisotropy of the network, there also exists a second shear modulus for shear applied at $60$ degrees to the $x-y$ shear, given by $\sqrt{3} (\alpha_m + \gamma_m)/8$, and can be obtained from the theory. This work focused on shear strains applied in the $x-y$ direction for the calculation of the shear modulus, and uniform expansion in all directions for the bulk modulus. 
 
The rigidity percolation thresholds from our EMT agree with Maxwell constraint counting with the threshold depending on both $p_x$ and $p_y$.   And unlike the isotropic case where $G$ and $K$ increase linearly with $p$ above the transition, the $G$ and $K$ versus $p_y$ curves (at fixed $p_x$) exhibit a slight departure from linearity. While this beyond-linear aspect is not as dramatic as in the EMT with spring networks with additional angular springs~\cite{heussinger06,das,broedersz,dascrosslinks,mao3}, anisotropy is yet another way to generate nonlinearity in the $G$ vs $p$  characteristics in a disordered solid, even at small strains. Although the stress-strain relationship continues to be linear at small strains studied in this work, we speculate that anisotropy would have a non-trivial contribution to the nonlinear scaling of the shear modulus, and the differential shear modulus as a function of strain at large strains, and will be studied in future work.

Drawing further comparison with the isotropic triangular lattice, we find in two limiting cases, discontinuous onsets of the bulk and/or shear modulus. Such discontinuous onsets do not occur in the isotropic case.  When $p_x=1$, the bulk modulus jumps discontinuously from zero to a finite value as a function of $p_y$, while the shear modulus increases continuously from zero. 
When $p_y=1$, both the bulk and shear modulus jump discontinuously from zero to a finite value at $p_x=0$. The latter result differs from recent work adding additional next-nearest-neighbor springs to a square lattice of springs. In this recent work, the shear modulus is equal to effective spring constant due to these additional springs and scales quadratically with the occupation probability of the next-nearest-neighbor bonds, i.e. it increases beyond zero continuously. In the anisotropic case, the ``additional'' $p_x$ bonds are nearest-neighbors bonds and the effective medium elastic constant scales linearly with $p_x$ (see inset of Fig.~\ref{Figure8}) after some initial non-zero value at $p_x=0$ due to the $x$-component of the $p_y$ bonds contributing to elasticity (in the $x$-direction).

We compare our EMT with lattice simulations and find rather good agreement, particularly for larger values of $p_x$ and $p_y$. As with lattice simulations in the isotropic case, the rigidity percolation threshold is lower than the EMT value. For reference, in the isotropic case, the EMT threshold value is $p_{rp}=2/3$, while simulations yield $p_{rp}=0.6602\pm0.0003$~\cite{thorpe2} the $p_{rp}$s to the infinite system limit in the isotropic case is difficult. For the anisotropic case, the task is further complicated by anisotropic finite-size scaling with different relevant length scales in the two directions such that one needs to rescale the $x$ and $y$ directions by different amounts.  Anisotropic finite-size scaling in, say, directed percolation, has been done and is based on a field theory able to estimate the two different length scales~\cite{directed}.  The absence of a field theory for rigidity percolation leaves one little to hang his/her hat on and so we leave this for future work.   

While our lattice simulations are not as in good agreement with the EMT as in the isotropic case, this discrepancy is due, in part, to finite-size effects, which tend to be more complex in anisotropic systems than isotropic systems given the presence of different lengthscales. On the other hand, one could argue that the rather good agreement is unexpected since the effective dimensionality of the system may differ from the isotropic case where mean field and two-dimensional predictions behave similarly.  

To our knowledge, our EMT calculation is the first for anisotropic spring networks and lays the foundation for the next stage where we will consider the presence of bending elasticity and non-linear response to more accurately mechanically model cytoskeletal filaments in lamellipodia where the actin filaments have a preferred orientation~\cite{svitkina,blanchoin}. We would then be able to better quantify the collective elastic response of a system that is important for cell motility on two-dimensional substrates. This will also allow for better comparison with recent work on the mechanics of anisotropic semiflexible polymer networks~\cite{levine,zippelius,head}. It would also be interesting to investigate the effects of splay rigidity~\cite{harris1,harris2} in anisotropic network models with bending, which may turn out to be more generic than originally thought and may be relevant for packing derived anisotropic networks based on granular materials and emerging liquid crystalline order in elastomeric gels. Such effects can be quantified with development of anisotropic effective medium theories going beyond the one constructed here. 
 
TZ and JMS acknowledge support from the Soft Matter Program at Syracuse University, and MD acknowledges support from the RIT College of Science via a D-RIG grant.  
JMS and MD also acknowledge the hospitality of the Aspen Center for Physics  (supported by NSF grant \#1066293) where some preliminary discussions on the project took place.

\end{document}